\documentclass[prl,twocolumn,showpacs,aps,superscriptaddress,floatfix]{revtex4}

\usepackage{graphicx}
\usepackage{bm}
\usepackage{dcolumn}
\usepackage[usenames]{color}
\begin{document}
%-------------------------------------------------------------------------------
\title{The nonlocal correlation: A key to the solution of the \em{CO adsorption
 puzzle}}

\author{P.~Lazi\'{c}\footnote{On leave of absence from the Rudjer
Bo\v{s}kovi\'{c} Institute, Zagreb, Croatia.}}
\affiliation{Institut f\"ur Festk\"orperforschung (IFF) and Institute for
Advanced Simulation (IAS), Forschungszentrum J\"ulich, 52425 J\"ulich,
Germany}
\author{M. Alaei}
\affiliation{Institut f\"ur Festk\"orperforschung (IFF) and Institute for
Advanced Simulation (IAS), Forschungszentrum J\"ulich, 52425 J\"ulich,
Germany}
\affiliation{Department of Physics, Isfahan University of
Technology, Isfahan 84156, Iran}
\author{N. Atodiresei}
\affiliation{Institut f\"ur Festk\"orperforschung (IFF) and Institute
for Advanced Simulation (IAS), Forschungszentrum J\"ulich, 52425
J\"ulich,
Germany}
\affiliation{The Institute of Scientific and Industrial Research, Osaka
University, 567-0047 Osaka, Japan}
\author{V.~Caciuc}
\affiliation{Physikalisches Institut, Westf\"alische Wilhelms
Universit\"at M\"unster, Wilhelm-Klemm-Str.~10, 48149 M\"unster, Germany}
\affiliation{Institut f\"ur Festk\"orperforschung (IFF) and Institute for
Advanced Simulation (IAS), Forschungszentrum J\"ulich, 52425 J\"ulich,
Germany}
\author{R.~Brako}
\affiliation{Rudjer Bo\v{s}kovi\'{c} Institute, Zagreb, Croatia}
\author{S.~Bl\"ugel}
\affiliation{Institut f\"ur Festk\"orperforschung (IFF) and Institute for
Advanced Simulation (IAS), Forschungszentrum J\"ulich, 52425 J\"ulich,
Germany}
\date{\today}

\begin{abstract}
We study the chemisorption of CO molecule into sites of different 
coordination  on (111) surfaces of late 4d and 5d transition metals. In an 
attempt to solve the well-known \textit{CO adsorption puzzle} we have
applied the relatively new vdW-DF theory of nonlocal correlation. 
The application
of the vdW-DF functional in all considered cases improves or completely
solves the discrepancies of the adsorption site preference and improves the value 
of the adsorption energy. By introducing a cutoff distance for nonlocal interaction 
we pinpoint the length
scale at which the correlation plays a major role in the systems considered.
\end{abstract}
%-------------------------------------------------------------------------------

\maketitle
%-------------------------------------------------------------------------------

For decades the density functional theory (DFT)~\cite{DFT} has been the
main theoretical tool used to analyze, understand and predict material properties
and chemical processes. Although very successful in explaining the finest
details of many systems, even very complex ones,
%~\cite{Wende}, 
it fails in
some cases.  In particular it fails in weakly bound systems where van der
Waals forces play a major role. Examples are numerous. Practically all
biological systems and any system containing noble gas atoms cannot
be calculated correctly within the current approximations to the DFT,
the local (LDA) and  the semilocal generalized gradient approximation
(GGA). A practical
solution to this kind of problems was relatively recently proposed
by Dion {\em et al.}~\cite{Dion}. They have developed a new exchange-correlation
functional named vdW-DF, which is parameter free and therefore is
%can be
considered \textit{ab initio}. The main idea of the vdW-DF functional
is the introduction of nonlocal correlation into DFT calculations. The
failure of local and semilocal functionals in weekly bounded systems is
expected, but what is puzzling is that current approximations to the DFT
fail to predict the correct adsorption site of CO molecules on several
(111) metal surfaces, of which the Pt is the most famous example. Adsorption
of CO on Pt surface is of tremendous practical importance, in particular
in exhaust catalyst in the car industry. Some eight years ago Feibelman
{\em et al.}~\cite{Feibelman} 
published a comprehensive study of CO on Pt(111), performing calculations
independently with several state-of-the-art DFT computer codes, 
for a variety of exchange-correlation potentials. 
%pointed out 
They found
that DFT systematically predicts that
CO prefers the hollow site on Pt(111) surface, while experiments show the
adsorption into the top, sometimes bridge but never hollow site. Due to
expected success of DFT in such chemisorbed systems the problem was named
the \textit{CO adsorption puzzle}. The same \textit{puzzle} appears on
some other (111) surfaces such as Cu, Rh, Ag, and Au. Even more confusing
is the correct prediction of the hollow site on some other surfaces such
as Pd(111) in agreement with experiment. Additionally to the correct site preference getting the correct barrier height 
between the sites of different coordination is of tremendous importance for the analysis of diffusion phenomena on 
surfaces as discussed by Hu et al. \cite{Hu}. By considering the \textit{puzzle} problem 
they have also noticed that the lack of nonlocal correlation is one of the possible sources of the problem.

In order to solve the \textit{puzzle} we apply the vdW-DF
functional. Even if up to now vdW-DF was mostly used to treat van der
Waals type of systems, achieving quite good success~\cite{Chakarova}, the
functional itself was constructed in a seamless fashion, i.e.\ describing
the nonlocal correlation correctly 
at all length scales.
%in all situations. 
In fact, the functional has recently been successfully applied to a
covalently bonded system \cite{Johnston}.
We apply the vdW-DF functional as a post-processing procedure on
a charge density obtained from a standard GGA calculation. Such an
application of the functional has been justified by the work of Thonhauser
{\em et al.}~\cite{Thonhauser}, who implemented the functional in a fully
selfconsistent manner into a DFT code and showed that 
%the change occurs
%only in the total energy of the system, while 
the change of Kohn-Sham
eigenstates was 
%found to be 
negligible.

Following the original procedure of Dion {\em et al.}~\cite{Dion} we replace
the conventional GGA correlation energy by a sum of two terms, the LDA
correlation and a newly defined nonlocal term $E_{\mathrm{ c}}^{\mathrm{ nl}}$:
\begin{equation}
\label{newtotal}
E_{\mathrm{ vdW\text{-}DF}}=E_{\mathrm{ GGA}}-E_{\mathrm{ GGA,c}}+E_{\mathrm{ LDA,c}}+E_{\mathrm{
c}}^{\mathrm{ nl}}
\end{equation}
where $E_{\mathrm{ GGA}}$ is the selfconsistent GGA total  energy.
The last term in Eq. (\ref{newtotal}) is completely determined by the
electronic charge density and requires the evaluation of a double space
integral: 
\begin{equation}
\label{nonlocal}
E_{\mathrm{c}}^{\mathrm{nl}}=\frac{1}{2} \int d^3 r \, d^3 r'\,  n(\textbf{\textit{r}})
\phi(\textbf{\textit{r}},\textbf{\textit{r}}')
n(\textbf{\textit{r}}')
\end{equation}
where $n(\textbf{\textit{r}})$ is the charge
density at point $\textbf{\textit{r}}$ and $\phi
(\textbf{\textit{r}},\textbf{\textit{r}}')$ is a kernel function described
in Ref.~\cite{Dion}.
Since to our knowledge there is no available code to calculate the vdW-DF
functional we have developed our own~\cite{JuNoLo}.
The code has been tested on the few examples which have been published 
so far~\cite{Dion}, achieving a complete agreement.
Our \textit{ab initio} DFT calculations were carried out mostly
by employing the pseudopotential plane-wave formulation
as implemented in PWscf~\cite{pwscf}. In the spirit of the original
\textit{puzzle} article~\cite{Feibelman} we performed calculations for the experimentally most
interesting coverage of 1/4 CO monolayer (ML) 
using two other codes, the Dacapo program~\cite{dacapo,dacapo2},
which is an ultrasoft pseudopotential plane-wave code similar to PWscf,
and VASP~\cite{vasp,VASP1,VASP2}, which is the most accurate 
of the three due to its PAW implementation~\cite{PAW}. 

It has been argued by the authors of the vdW-DF functional and others
\cite{Dion,Johnston} that the vdW-DF should be used together with the
revPBE functional for the calculation of the exchange energy contribution,
because it seems to approximate most correctly the exact exchange and
by that avoids some spurious binding effects. We have used both PBE
\cite{PBE} and revPBE~\cite{revPBE} functionals in our DFT calculations.

\begin{table*}[ht]
\centering
\begin{centering}
%\scriptsize      %%%% For preprint
\begin{tabular}{|c c|r r r|r| r r r c|}
\hline
  &  & &$E^{\mathrm{ top}}-E^{\mathrm{ hollow}} $ &[meV]                &
  [meV] &&  $E_{\mathrm{ ads}}^{\mathrm{ top}}$ & [meV] &                 \\
  
$\theta$ [ML] &system       & GGA   & PBE+NL &revPBE+NL& NL shift
& GGA      & PBE+NL  & revPBE+NL& Exp. Ref.~\cite{Pedersen} \\
\hline 
              &P Pt      & 198   & 95    & 35*     &  $-$103
              & $-$1648        & $-$1442       & $-$1185*   & top
              \\ 
             &Pr Pt   & 93    & 56*    &  $-$3   &  $-$96
             & $-$1331        & $-$1390*      & $-$1135    & --''--
             \\
$\frac{1}{12}$ &P Cu      & 170   & 50     & $-$14*  &  $-$120    &
$-$606        & $-$450       & $-$210*             & --''--         \\ 
              &Pr Cu   & 53     & 102*     &    13    &  $-$40
              & $-$373        & $-$594*      & $-$290         & --''--
              \\
              &P Rh      & 49    & $-$101 & $-$166* &  $-$150    &
              --        & --       & --           & --''--         \\ 
              &P Pd      & 631   & 526    & 454*    &  $-$105    &
              --        & --       & --            & fcc hollow      \\ 
\hline
              &P Pt      & 170   & 120    & 58*     &   $-$50     &
              $-$1630 & $-$1460 & $-$1212* & $-$1370$\pm$130 top    \\ 
              &V Pt      & 100   & 56     &  1*     &  
              $-$44     & $-$1572  & $-$1373 & $-$1124*    & --''--
              \\
              &D Pt      & 220   & 152    & 94*     &  
              $-$68     & $-$1616  & $-$1512 & $-$1252* & --''--
              \\  
              &Pr Pt   & 60    & 81*    & 20      &   $-$40
              & $-$1326  & $-$1404*      & $-$1160  & --''--
              \\ 
$\frac{1}{4}$  &P Cu      & 150   & 50     & $-$43*  &   $-$100
&  $-$800  & $-$720  & $-$492*  &  $-$500$\pm$50 top     \\
              &V Cu      & 139   & 34     &  $-$32* &  
              $-$105    & $-$679   & $-$620  & $-$366*  & --''--
              \\
              &Pr Cu   & 20     & 30*     & $-$30       &   $-$50      & $-$540 
              & $-$843*      & $-$520        & --''--                    \\ 
              &P Rh      & 30    & $-$100 & $-$168* &   $-$130    &
              $-$1890 & $-$1840 & $-$1620* &  $-$1450$\pm$140 top   \\ 
\hline
\end{tabular}
\caption{ 
Comparison of the energies of CO adsorption on (111) surfaces of several
metals, calculated with the standard GGA and the vdW-DF post-processing
approach. The abbreviation in the first column denotes the program used:
V--VASP, D--Dacapo, P--PWscf. The letter ``r'' refers to the selfconsistent
use of the revPBE functional, while otherwise PBE (VASP, PWscf)
or PW91~\cite{PW91} (Dacapo) were used. Columns 2--4: The energy 
differences between the top and the hollow site. The results marked by 
an asterisk are obtained using an exchange functional different from the 
one with which the electronic density was calculated. Column 5: The 
`NL shift', i.e.\ the change between the values in column 2 and column 
3 or 4 (whichever corresponds to the selfconsistent functional).
Columns 6--8: The calculated adsorption energies for the top site, 
column 7: The experimental value, where available. In all cases CO
molecule is at GGA equilibrium position.
}
\label{table1}
\end{centering}
\end{table*}

We have applied the vdW-DF functional to the adsorption of CO into top
and fcc hollow sites on a whole range of (111) surfaces which have been
related to the CO puzzle, for two different coverages. 
The results are summarized in Table~\ref{table1}. 
We immediately notice the important result that in almost all cases the 
energy of adsorption into the top site is lowered by roughly 100 meV 
compared to that of the hollow site. In the case of Rh this
completely solves the \textit{puzzle}, while for Pd the hollow site 
remains more stable, which is in agreement with experiment.
The shift in energy is somewhat smaller for Pt and the 
preference for the hollow site remains, confirming its reputation as
the most difficult \textit{puzzle} system.  Furthermore, in all cases 
the adsorption energy is improved compared to the pure GGA value.

\begin{figure}[htb]
{
\resizebox{0.90\columnwidth}{!}{
\includegraphics[clip=true]{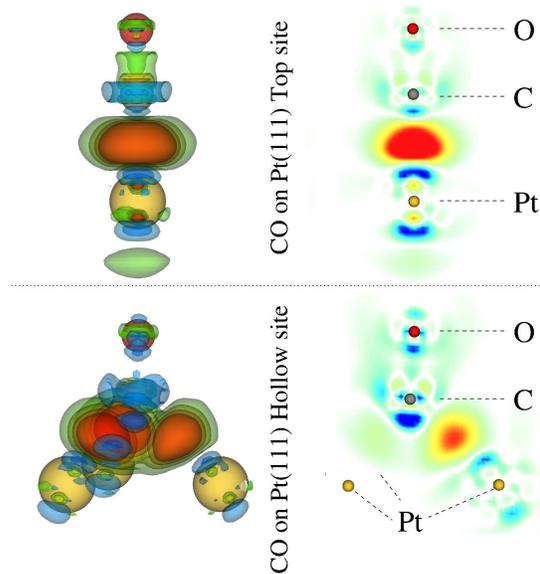}
}
}
\caption{Differences in correlation energy
densities between revPBE ($E_{\mathrm{ revPBE,c}}$) and vdW-DF ($E_{\mathrm{
LDA,c}}+E_\mathrm{c}^{\mathrm{ nl}}$) calculations, for top site, upper
panel, and hollow site, lower panel. 
The three-dimensional representation (left) and the cut along the CO molecule 
and through (the nearest) metal atom of the Pt surface (right) are shown.
Red (blue) colors show regions in which PBE yields larger (smaller)
binding energies compared to the vdW-DF functional. 
Color scales are the same
in both images, with green representing zero value. 
By summing 
all the contributions 
we found out that PBE correlation overbinds 
compared to vdW-DF in both cases.
To get insight into this figure please follow the links: \\
http://www.youtube.com/watch?v=IbPhmFb-Vsg \\ 
http://www.youtube.com/watch?v=rqxwZOMaWbg
}
\label{pbeoverbind}
\end{figure}

In order to understand the difference between the PBE (or any GGA) 
and the vdW-DF correlation energy 
we have examined the spatial distribution of the correlation energy densities. For both  we have 
calculated the distribution for the CO chemisorbed on metal, and
subtracted the values for the clean surface and an isolated molecule
(at exactly the same positions as in the coupled system), in order to find
in which regions of space changes occur upon chemisorption.
Since the PBE correlation depends upon the gradient of the charge density
and the vdW-DF correlation upon the charge density itself through the 
integral~(\ref{nonlocal}), their spatial distribution is quite different. 
In Fig.~\ref{pbeoverbind} we show the difference of the two.
In the red regions the PBE correlation overbinds compared to vdW-DF,
while the opposite is true in the blue regions. Integration over space
shows that the total effect is always an overbinding by PBE, but it is 
more pronounced in the fcc hollow case than in the top case, by about
100~meV, as discussed in Table~\ref{table1}. The figure suggests that 
this is due to the fact that there are three regions of overbinding
in the fcc hollow case and only one in the top case (although the
latter is somewhat stronger than any single hollow one).
 
%

%Our conclusion is that PBE correlation is actually qualitatively quite good, but
%in some systems the quantitative error described above adds up in such a manner
%that it makes PBE look qualitatively wrong, e.g.\ due to the failure to correctly 
%predict adsorption sites in seemingly simple systems---this was the \textit{puzzle}.
We conclude that the PBE correlation functional makes a quantitative error
in evaluating correlation energies of different adsorption sites in these
systems and therefore seems to yield wrong results, e.g.\ due to the failure to
correctly predict adsorption sites in seemingly chemically bonded systems---this
was the \textit{puzzle}.
PBE otherwise correctly captures the physics of the problem 
as our relaxation
tests for \textit{puzzle} systems of Pt and Cu show. 
Since we do not have access to forces which would appear in a
selfconsistent implementation of the vdW-DF theory, we have relaxed the CO molecule by
hand. We first changed the position of the molecule relative to the surface in
the perpendicular direction while keeping the C--O distance at 
the relaxed value of the GGA calculation, 
and later we also changed that distance. It turns out
that the position of CO determined in GGA calculations is the optimal one
also for the vdW-DF functional. 
Moreover, the curvature of the
adsorption energy around the minimum does not change if we apply
vdW-DF theory although the value of the minimum shifts. 
In systems in which the PBE does not describe
the physics correctly, as for example a Kr dimer, both parameters---the
position of energy minimum and the curvature around it---are strongly changed
upon the application of the vdW-DF theory~\cite{Brako}.

We have also made some calculations of CO adsorption on (111) surfaces
of noble metals Au and Ag, and found that the results are qualitatively
different to the other considered surfaces. The 
experimental chemisorption energies
of CO on these surfaces are in the range of 300--400~meV. 
In comparison, pure GGA  energies are somewhat too small, which is
more reminiscent of true
van der Waals systems where GGA clearly underbindes, than  of the
strongly bonded systems
considered in this paper.
Also, while
for other systems the equilibrium position remains the same after
applying the vdW-DF functional, we found that for Au and Ag the
position of the minimum moves outwards, and (in agreement with experiment) 
the top site becomes clearly preferred. These properties single out
the noble metals as rather different from the other cases considered,
and we leave the detailed study for the future.

In order to further investigate how the nonlocality in the vdW-DF
correlation compares with the usual semilocal GGA approach
we have introduced a cutoff distance in
our calculation of Eq. (\ref{nonlocal}) by a function
\begin{equation}
\label{eqdcut}
\theta(d_{\mathrm{cut}}-|\textbf{\textit{r}}-\textbf{\textit{r}}'|)
\end{equation}
under the integral. 
The cutoff $d_{\mathrm{cut}}$ defines the largest allowed distance
of the nonlocal correlation.

\begin{figure}[htb]
\resizebox{0.97\columnwidth}{!}{
{\includegraphics[clip=true]{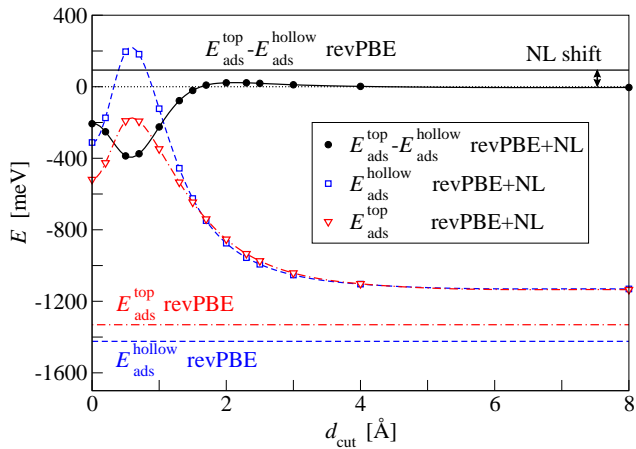}} 
}
\caption{The energies of adsorption of a CO molecule at top 
and hollow sites on the Pt(111) surface at a coverage of $1/12$, and the 
difference of the two. The calculation was done 
using the density obtained in a revPBE GGA calculation, 
varying the cut-off 
in
Eq.~(\ref{eqdcut}).
At $d_{\mathrm{cut}}=0$ (i.e.\ including GGA revPBE exchange but only the
LDA term in correlation)
the top site is clearly preferred. 
The full revPBE+NL
(large $d_{\mathrm{cut}}$) 
gives a small preference for the top site, while the plain
GGA shows a strong preference for the hollow site, at variance with experiment.
}
\label{dcut}
\end{figure}

We performed calculations 
for CO chemisorbed on Pt(111) in top and hollow site
at equilibrium distance, and subtracted the energy for the separated molecule and
surface, all calculated with the same value of $d_{\mathrm{cut}}$. The results are 
shown in Fig.~\ref{dcut}. Three regimes are immediately apparent. At small
$d_{\mathrm{cut}}$, less than $\sim 1$~{\AA}, the contribution to the 
nonlocal correlation comes from within regions where charge is accumulated
or depleted upon chemisorption (roughly corresponding to regions of intense colors in Fig.~\ref{pbeoverbind}). 
This region is somewhat unphysical, because in the integral~(\ref{nonlocal}) only a small part of the 
kernel $\phi$ contributes.
In the region $\sim 2-3$~{\AA},
which is the characteristic length of the chemical bond (and of the distances
between blue blobs in Fig.~\ref{pbeoverbind}) the nonlocal correlation 
becomes more binding. 
These changes occur faster for the hollow than for the top site, reflecting 
the fact that there are three bonding regions instead of one.
Finally, in the true van der Waals region, say
for $d_{\mathrm{cut}} > 4$~{\AA}, the interaction is, of course, attractive, 
but this region contributes little to the strongly chemisorbed system of 
CO on Pt(111), and almost nothing to the difference between the top and
hollow site.

Fig.~\ref{dcut} shows that the 
physical length scale of the nonlocal correlation energy, i.e.\ when the energies and the energy differences saturate,  
is of the order of the distance between the
carbon and the substrate atom. 
However, at this length scale the exact geometry of the chemisorbed complex 
becomes relevant, which intuitively explains why the GGA,
which depends only upon the local value of the density gradient and thus
does not see the big picture, is not able to give the correct 
values,
in particular the energy differences between sites of different coordination.

% added for arxiv

\begin{figure}[htb]
{
\resizebox{0.95\columnwidth}{!}{ \includegraphics[clip=true]{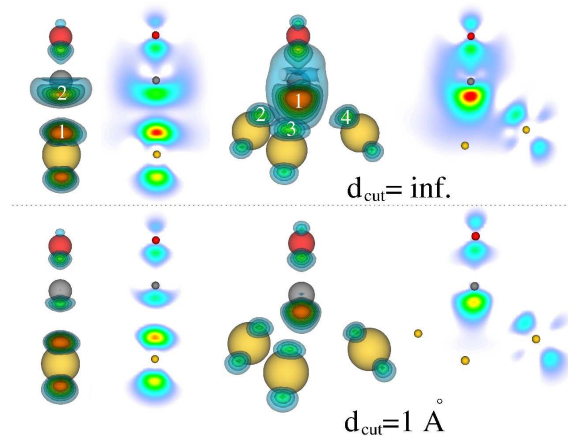}}
}
\caption{(Color online) Binding energies due to NL term calculated with infinite $d_{\rm cut}$, upper panel, and with $d_{\rm cut}$=1 ${\rm \AA}$ lower panel. Left (right) column shows top (hollow) site adsorption. Color scales are the same in all four images, blue color represents zero value. On the lower right image the sphere of diameter of 3 $\rm \AA$ is shown while the spheres representing Pt atoms have 1 $\rm \AA$ in diameter. From the upper images it is clear that the most of the binding energy is coming from two regions in the top and four regions in the hollow case which are numerated in the figure. What remains unclear from the upper panel is whether the correlation binding energy in marked regions is due to intra-region or inter-region interaction. (Comparing these images to the ones in Fig.~\ref{pbeoverbind} one can see that these regions correspond to the ones where PBE shows underbinding compared to vdW-DF, and that they are actually regions where the charge density was depleted due to binding). To answer this question we have performed calculations with cutoff parameter value $d_{\rm cut}$=1 ${\rm \AA}$ which is just large enough to take all intra-region interaction into account, while avoiding any inter-region interactions (for the length scale reference in the figure the spheres are drawn). If there is no inter-region interactions the corresponding images from the upper and lower panel would not differ. However, a large difference between images exists, proving the presence of strong inter-region interactions. To get insight into this figure please follow the links: \\
http://www.youtube.com/watch?v=kgMNEulddoo \\
http://www.youtube.com/watch?v=6iuFD81hucY 
}
\label{nonlocalcut}
\end{figure}

In order to understand more thoroughly how nonlocal correlation binding energy depends on $d_{\rm cut}$ we have visualized the correlation energy density for two different values of parameter $d_{\rm cut}$, for top and hollow site adsorption, as shown in Fig.~\ref{nonlocalcut}.

From Fig.~\ref{nonlocalcut} we conclude that both intra- and inter-region contributions exist. Comparing Fig.~\ref{dcut} and Fig.~\ref{nonlocalcut} one can see that inter-region interaction starting to take place at $d_{\rm cut} \approx 2 \rm \AA$ is the most important nonlocal contribution to the energy difference between top and hollow site, and that it is also responsible for the change of the adsorption energy. Looking carefully in the upper panel of Fig.~\ref{nonlocalcut} it is noticeable that most of the nonlocal correlation binding energy in the top site adsorption is located just above the Pt metal atom (region 1), while in the hollow site case most of it is found just underneath the carbon atom (region 1), reflecting different chemical bonding mechanism in the two cases.

% end of addition

The fact that the PBE functional as well as other GGA functionals overbind
the adsorption of molecules on metal surfaces was realized some time
ago~\cite{Hammer}. 
In the
whole letter we made little reference to the exchange energy, but  followed
the arguments of the authors of the vdW-DF theory, and considered the
revPBE formulation as a possible improvement.
However, we believe that the
implementation of the exact exchange might be the best approach,
and may even lead to the resolution of the \textit{puzzle}.

In conclusion, we have shown that the application of vdW-DF theory resolves
a large part of the \textit{CO adsorption puzzle}. 
Even though in almost all cases we achieve a preference of the top site, we cannot claim a complete 
solution of the puzzle, because larger energy differences are required to agree with experimental thermal stability.
We
have shown that nonlocal correlation effects at a typical length scale
of C--metal atom distance ($\sim 2-3$~{\AA}) are decisive for a
good description of CO \textit{puzzle} systems. 
We are not sure whether the construction of a better semilocal functional is
impossible, as was recently concluded on the grounds of exploration of
the \textit{puzzle} problem~\cite{Stroppa}, but 
in the construction of such a semilocal functional (which would be
desirable due to a lower computational cost) one should take as a guide
and the ultimate check of the quality of the functional
the treatment of the same CO \textit{puzzle} systems 
with vdW-DF,  
as it reveals the delicate balance between nonlocal correlation and
adsorption site preference.
%In this letter we have avoided discussions in terms of
%position of d-bands and hybridization of states, which is common
%in 
%literature about the \textit{puzzle} problem. 
%The reason
%is that we do not implement vdW-DF theory selfconsistently
%but rather as a post-processing technique as justified by Thonhauser et
%al.~\cite{Thonhauser},
%which is, however, based on a
%calculation for a particular example system.
%It is not clear that
%in the case of CO \textit{puzzle} systems this conclusion need not be
%reconsidered, 
%and whether a
%selfconsistent implementation of the vdW-DF theory 
%may even lead to significant further improvements.
To summarize, our calculations show that the vdW-DF theory 
is crucial in chemically bonded systems
which were initially not expected to be much affected by it.

The calculations were performed on the JUMP and Blue/Gene
supercomputers at the Forschungszentrum J\"ulich, Germany.
Two of us, P.~L.\ and S.~B., thank the Deutsche Forschungsgemeinschaft (DFG)
(Priority Programme ``Molecular Magnetism'') and Alexander von Humboldt foundation for financial support. 
N.~A. acknowledges the support of Japan Society for the Promotion of Science.
R.~B. acknowledges the support of MSES of the Republic of Croatia through project No. 098-0352828-2836.

\end{document}